\documentclass[conference,final]{IEEEtran}
\usepackage{mathptmx}

\def\hb{\hbox to 10.7 cm{}}

\usepackage{booktabs} 
\usepackage{listings}
\usepackage{fancyvrb}
\usepackage{wrapfig}
\usepackage{graphicx}
\usepackage{amsmath}
\usepackage{amssymb}
\usepackage{color}
\usepackage{ifpdf}
\usepackage{subcaption}
\usepackage{float}
\usepackage[utf8]{inputenc}

\usepackage{multirow}
\usepackage{rotating}
\usepackage{setspace}
\usepackage{amsmath}
\usepackage{moresize}
\usepackage{url}
\usepackage{booktabs}
\usepackage{listings}
\usepackage{paralist}
\usepackage{wrapfig}
\usepackage{multirow}
\usepackage{ifpdf}
\usepackage{xspace}
\usepackage{keyval}
\usepackage{color}
\usepackage[font=small,labelfont=bf]{caption}

\definecolor{listinggray}{gray}{0.95}
\definecolor{darkgray}{gray}{0.7}
\definecolor{commentgreen}{rgb}{0, 0.4, 0}
\definecolor{darkblue}{rgb}{0, 0, 0.4}
\definecolor{middleblue}{rgb}{0, 0, 0.7}
\definecolor{darkred}{rgb}{0.4, 0, 0}
\definecolor{brown}{rgb}{0.5, 0.5, 0}

\usepackage[normalem]{ulem}
\makeatletter
\def\cyanuwave{\bgroup \markoverwith{\lower3.5\p@\hbox{\sixly \textcolor{cyan}{\char58}}}\ULon}
\def\reduwave{\bgroup \markoverwith{\lower3.5\p@\hbox{\sixly \textcolor{red}{\char58}}}\ULon}
\def\blueuwave{\bgroup \markoverwith{\lower3.5\p@\hbox{\sixly \textcolor{blue}{\char58}}}\ULon}
\font\sixly=lasy6 
\makeatother

\usepackage{xcolor}
\usepackage[colorlinks]{hyperref}
\AtBeginDocument{%
  \hypersetup{
    citecolor=blue,
    linkcolor=blue,   
    urlcolor=blue}}

\newif\ifdraft
\ifdraft
\definecolor{ocolor}{rgb}{1,0,0.4}
\newcommand{\onote}[1]{ {\textcolor{ocolor} { (***Ole: #1) }}}
\newcommand{\terminology}[1]{ {\textcolor{red} {(Terminology used: \textbf{#1}) }}}

\newcommand{\jhanote}[1]{ {\textcolor{red} { ***shantenu: #1 }}}
\newcommand{\alnote}[1]{ {\textcolor{blue} { ***andreL: #1 }}}
\newcommand{\amnote}[1]{ {\textcolor{blue} { ***andreM: #1 }}}
\newcommand{\georgenote}[1]{ {\textcolor{brown} { ***sharath: #1 }}}
\newcommand{\revThreeNote}[1]{ {\textcolor{purple}{}}} 
\newcommand{\revOneNote}[1]{ {\textcolor{purple}{}}} 
\newcommand{\revTwoNote}[1]{ {\textcolor{purple}{}}} 
\definecolor{orange}{rgb}{1,.5,0}
\newcommand{\aznote}[1]{ {\textcolor{orange} { ***ashley: #1 }}}
\definecolor{dandelion}{cmyk}{0,0.29,0.84,0}
\newcommand{\mtnote}[1]{ {\textcolor{dandelion} { ***matteo: #1 }}}
\newcommand{\note}[1]{ {\textcolor{magenta} { ***Note: #1 }}}
\else
\newcommand{\onote}[1]{}
\newcommand{\terminology}[1]{}

\newcommand{\alnote}[1]{}
\newcommand{\amnote}[1]{}
\newcommand{\athotanote}[1]{}
\newcommand{\georgenote}[1]{}
\newcommand{\pmnote}[1]{}
\newcommand{\jhanote}[1]{}
\newcommand{\msnote}[1]{}
\newcommand{\mrnote}[1]{}
\newcommand{\aznote}[1]{}
\newcommand{\mtnote}[1]{}
\newcommand{\note}[1]{}
\newcommand{\revOneNote}[1]{}
\newcommand{\revTwoNote}[1]{}
\newcommand{\revThreeNote}[1]{} 
\fi

\newcommand{\pilot}{Pilot\xspace}

\newcommand{\pilotjob}{pilot-job\xspace}

\newcommand{\pilotdescription}{Pilot-Description\xspace}

\newcommand{\computeunit}{compute-unit\xspace}
\newcommand{\computeunits}{compute-units\xspace}

\newcommand{\upp}{\vspace*{-0.5em}}

\lstdefinestyle{myListing}{
  frame=single,
  backgroundcolor=\color{listinggray},
  language=C,
  basicstyle=\ttfamily \footnotesize,
  breakautoindent=true,
  breaklines=true
  tabsize=2,
  captionpos=b,
  aboveskip=0em,
  belowskip=-2em,
}

\lstdefinestyle{myPythonListing}{
  frame=single,
  backgroundcolor=\color{listinggray},
  language=Python,
  basicstyle=\ttfamily \scriptsize,
  breakautoindent=true,
  breaklines=true
  tabsize=2,
  captionpos=b,
}

\ifpdf
\DeclareGraphicsExtensions{.pdf, .jpg, .tif}
\else
\DeclareGraphicsExtensions{.ps,  .eps, .jpg}
\fi

\usepackage{listings}
\usepackage{paralist}

\lstnewenvironment{code}[1][]%
{
\noindent
\minipage{1.0 \linewidth}
\vspace{0.5\baselineskip}
\lstset{
    language=Python,
    frame=single,
    captionpos=b,
    stringstyle=\ttfamily,
    basicstyle=\scriptsize\ttfamily,
    showstringspaces=false,#1}
}
{\endminipage}

\defaultleftmargin{1em}{}{}{}
\begin{document}




\title{Performance Characterization and Modeling of Serverless and HPC
Streaming Applications}

\author{Andre Luckow$^{1,2,3,4}$, Shantenu Jha$^{1,5}$\\
   {\footnotesize{\emph{$^{1}$RADICAL, ECE, Rutgers University, Piscataway,NJ 08854, USA}}}\\
   \footnotesize{\emph{$^{2}$BMW Group, Munich, Germany}}\\
   \footnotesize{\emph{$^{4}$Ludwig Maximilian University, Munich, Germany}}\\
   \footnotesize{\emph{$^{3}$Clemson University, Clemson, SC 29634, USA}}\\
   \footnotesize{\emph{$^{5}$Brookhaven National Laboratory, Upton, NY, USA}\upp\upp\upp}
}

\date{}
\maketitle

\begin{abstract}
Experiment-in-the-Loop Computing (EILC) requires support for numerous types of
processing and the management of heterogeneous infrastructure over a dynamic
range of scales: from the edge to the cloud and HPC, and intermediate
resources. Serverless is an emerging service that combines high-level
middleware services, such as distributed execution engines for managing tasks,
with low-level infrastructure. It offers the potential of usability and
scalability, but adds to the complexity of managing heterogeneous and dynamic
resources. In response, we extend Pilot-Streaming to support serverless
platforms. Pilot-Streaming provides a unified abstraction for resource
management for HPC, cloud, and serverless, and allocates resource containers
independent of the application workload removing the need to write
resource-specific code. Understanding of the performance and scaling
characteristics of streaming applications and infrastructure presents another
challenge for EILC. StreamInsight provides insight into the performance of
streaming applications and infrastructure, their selection, configuration and
scaling behavior. Underlying StreamInsight is the universal scalability law,
which permits the accurate quantification of scalability properties of
streaming applications. Using experiments on HPC and AWS Lambda, we
demonstrate that StreamInsight provides an accurate model for a variety of
application characteristics, e.\,g., machine learning model sizes and resource
configurations.

\end{abstract}





\section{Introduction}


The integration of experimental instruments, the Internet-of-Things (IoT), 
compute and data infrastructure -- from the edge to HPC and
cloud -- is increasingly important to enable new scientific discoveries and
advances~\cite{reed2019}. Experiment-in-the-Loop Computing (EILC) describes a
new type of application and infrastructure, where experimental capabilities are
augmented with compute and data capabilities to enable more intelligent
experiments. Stream processing capabilities are essential for EILC to enable
realtime insights on data feeds from myriad sources~\cite{streaming2015}.
Applications for stream processing range from connected and autonomous
vehicles~\cite{7938385}, realtime analysis of astronomy data to other
scientific experiments, such as light sources~\cite{pilot-streaming}. For
example, Synchrotron light source experiments, such as those at the National
Synchrotron Light Sources II (NSLS-II)~\cite{nsls} or the X-Ray Free Electron
Laser (XFEL) light sources. EILC and stream processing can be used for
detecting events of interests, pre-processing data, and steering of simulations 
and instruments. 



\alnote{add some more science relevance for streaming and serverless, e.g.
Maciej Malawski. 2016. Towards Serverless Execution of Scientific Workflows- HyperFlow Case Study.. In Works@ Sc. 25–33.
BlessonVarghese,PhilippLeitner,SuprioRay,KyleChard,AdamBarker,etal.
2019. Cloud Futurology. Computer (2019).
}

EILC require infrastructure --- edge, cloud, and HPC, which are complementary
to each other: edge is suited for realtime processing close to the data
source, HPC is better suited for high-end computational and data-intensive
tasks, such as simulations. Cloud platforms provide commoditized capabilities
for streaming and data analytics.

A new, emerging platform is \emph{serverless} --- which abstract away most,
but not all resource management concerns, such as the allocation of nodes,
containers and the management of the processing
framework~\cite{DBLP:journals/corr/BaldiniCCCFIMMR17, berkeley-serverless}.
Serverless is well suited for data-parallel tasks, including streaming.



EILC applications need to be able to orchestrate streaming workloads
comprising of dependent and parallel tasks. Managing such workloads on
heterogeneous infrastructure from HPC to serverless is associated with several
challenges:

\emph{Development \& Deployment:} There are a lack of abstractions for
efficiently managing task-based workloads on serverless, such as workloads
requiring complex inter-task dependencies and
parallelism~\cite{Baldini:2017:STF:3133850.3133855}. As a result, application
codes need to be carefully and manually partitioned and wrapped into
serverless APIs. While serverless promises to reduce the need for manual
resource management, it still requires applications to configure and adjust
resource-related parameters, e.\,g., the concurrency and memory per container.

\emph{Interoperability:} Abstractions for resource and task management differ
between HPC and serverless; HPC abstractions are low-level and
infrastructure-centric compared to serverless abstractions. Additionally,
differences between platform providers exist.  This lack of interoperability
complicates the use of multiple platforms within an application.



\emph{Execution:}  Multiple layers of infrastructure, middleware, and
execution engines make it difficult to monitor, understand and predict the
performance of an application. For serverless, many details of the runtime are
abstracted and opaque further complicating the performance understanding. The
determination of appropriate performance metrics and the manual evaluation of
streaming systems under different workloads and resource configurations is a
complex undertaking.



To overcome the development, interoperability, and execution challenges, we
extended \emph{Pilot-Streaming} to \emph{serverless} platforms.
\emph{Pilot-Streaming }is a unifying abstraction based on the pilot
abstraction~\cite{pstar12} for resource management across heterogeneous
platforms. The pilot abstraction decouples resource allocation from workload
execution, enabling scalable and flexible execution of tasks. Tasks are
self-contained sets of operations that possess different characteristics,
e.\,g., related to its granularity, compute, memory and I/O demands.
Data-parallel and streaming workloads are typically comprised of many
short-running tasks which in the case of streaming are generated in response
to incoming data. With the extension to serverless, Pilot-Streaming provides a
unified way to acquire HPC, clouds and serverless resources enabling
developers to compose tasks and express task-level parallelism efficiently.

We developed the \emph{StreamInsight} framework for analyzing and predicting
the performance of streaming systems and applications. It is based on the
Universal Scalability Law (USL)~\cite{Gunther1993ASC}, which captures the
underlying properties of the system that limit scalability. StreamInsight can
enhance the experimental design, automation and analysis of streaming systems.
We demonstrate  its capabilities using a machine learning streaming
application to detecting abnormal behavior on different infrastructures, in
particular on serverless using Kinesis~\cite{kinesis} as a message broker and
AWS Lambda~\cite{lambda} for processing, and on HPC machines using
Kafka~\cite{kreps2011kafka} as broker and Dask~\cite{dask} for processing.




This paper is structured as follows: In section~\ref{sec:background_related}
we provide essential background and related work. We continue with a
discussion of Pilot-Streaming on serverless in
section~\ref{sec:pilot-serverless}. In section~\ref{sec:perf_model}, we
present the theoretical foundation and architecture of StreamInsight. Further,
we validate the system by performing different experiments on different
infrastructures and application characteristics. We conclude with a discussion
of the results and future work in section~\ref{sec:conclusion}.

\section{Background and Related Work}
\label{sec:background_related}

In this section, we present important background and related work on
serverless  (section~\ref{sec:serverless}), and  benchmarking and
performance modeling of streaming applications (section~\ref{sec:benchmark}). 

\subsection{Serverless}
\label{sec:serverless}

\jhanote{Serverless or serverless computing?}\alnote{What do you think would be better suited? Just serverless? In addition to Faas there are other serverless services also for data and not just computing (e.g. athena)}

The term \emph{serverless}~\cite{berkeley-serverless} is most commonly used for
\emph{Function-as-a-Service (FaaS)} platforms, such as AWS
Lambda~\cite{lambda}, Google Cloud Functions and Azure Functions. It is a
computing paradigm that allows the scalable and fault-tolerant execution of
functions in response to defined events without the need to consider low-level
concerns, such as resource management, provisioning and scaling with respect
to the number of events. A serverless function comprises of a self-contained 
piece of code that implements an interface defined by the platform. It is 
then deployed on the FaaS infrastructure, which automatically 
instantiates, executes, scales and monitors the function.

Serverless  tightly integrates the abstraction for expressing computational
tasks with its instantiation in the runtime system. The function code is
required to be stateless, i.\,e., the state needs to be managed and stored
outside the function. Further, the function is subject to strict 
constraints, e.\,g., the container size is limited to a single core, a defined 
amount of memory, and the runtime is subject to a walltime (currently 15 
minutes processing time per event for Lambda). The cloud platform provides a
high-level SLA concerning the available compute and data resources.
Serverless  is increasingly relevant for scientific applications,
especially in conjunction with HPC
capabilities~\cite{DBLP:journals/corr/abs-1708-08028,JonasVSR17, numpywren}.
The event-driven nature of streaming applications makes them ideally suited to
utilize the serverless paradigm. \jhanote{serverless or serverless computing
paradigm?}\alnote{I removed serverless computing. Although in my
classification: serverless is broader than just computing, but in this paper
we are focusing on a particular serverless computing service}

\jhanote{I don't see mention of AWS GreenGrass?}\alnote{added to future work in sec V}

\subsection{Streaming Performance and Modeling}
\label{sec:benchmark}


While there is comprehensive work on benchmarking streaming
systems~\cite{Arasu:2004:LRS:1316689.1316732, yahoo-streaming}, typically,
these benchmarks do not reflect the complex and highly dynamic requirements of
real-world scientific streaming applications. Fox et al. propose the concept
of Big Data Ogres to describe well-understood application characteristics,
which served as a basis for the development of a set of benchmarks that
encapsulate commonalities of these applications~\cite{fox_bigdata_benchmarks}.
The concepts of Mini-Apps was introduced in the domain of data-intensive HPC
apps by Sukumar et al.~\cite{7840756}. Luckow et\,al. refined this concept for
streaming apps~\cite{pilot-streaming}.

Increasingly statistical methods are used to understand performance and to
make predictions, e.\,g., for resource (re)-configurations
decisions~\cite{fox_learning_everywhere}. Kremer-Herman et
al.~\cite{Kremer-Herman:2018:LMR:3291656.3291708} propose a model for
recommending the optimal infrastructure configuration for master/worker
applications. Ernest~\cite{ernest} is a system that combines analytical
modeling and data derived from a small set of performance counters and sample
runs. Cherrypick~\cite{cherrypick} uses a black-box model based on Bayesian
optimization to find optimal resource configurations for Big data frameworks
in the cloud. Caladrius~\cite{kalimcaladrius} is a streaming forecasting
system und provides a predictive model for traffic and an analytical model for
the processing system.

The main limitations of these approaches are that they only investigate a small
part of the stream processing pipeline and are often constrained to specific
frameworks. The experimental validation is often done using simple examples
that are not representative of real-world scientific applications. For
example, Caladrius is solely evaluated using a wordcount example and 
parallelism of 8.

\section{Pilot-Streaming: Extension to Serverless}
\label{sec:pilot-serverless}
\label{sec:pilotserverless}

\begin{figure}[t]
\centering
    \includegraphics[width=0.38\textwidth]{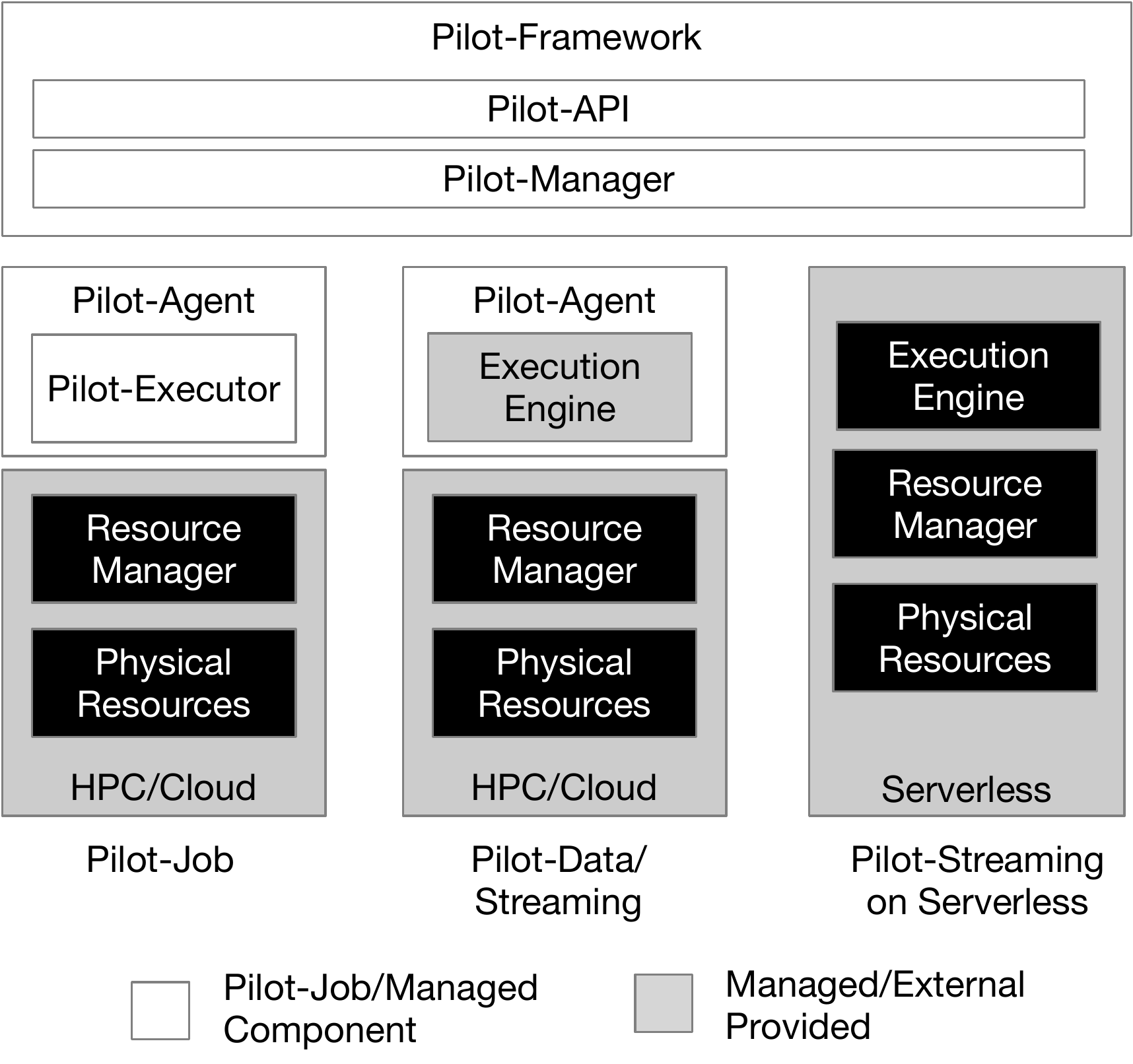}
  \caption{\textbf{Pilot Abstraction on Heterogeneous Platforms: HPC,
  Cloud, and Serverless:} The pilot abstraction can integrate with
  platforms and infrastructures at various levels. Pilot-Streaming on serverless
  abstracts the serverless-specific resource acquisition 
  enabling a unified way to manage tasks on HPC, clouds, and
  serverless. \jhanote{Is the Pilot-Job for HPC? And Pilot-Data for
  Cloud? Then maybe replace infrastructure at bottom of grey boxes with HPC
  and Cloud?} \alnote{That's where things get complicated: Pilot-Jobs can also run on EC2 VMs in the Cloud. Pilot-Data can support iRods on HTC, Lustre on HPC and S3 on clouds. I do agree that the primary mode would be Pilot-Job/HPC, Pilot-Data/Cloud....}\upp\upp\upp}
  \label{fig:figures_pilotserverless}
\end{figure}


In this section, we discuss the extension of \emph{Pilot-Streaming} to support
serverless platforms. Pilot-Streaming~\cite{pilot-streaming} provides a
unified interface for resource management on serverless and HPC, and supports
the development, deployment, and execution of streaming applications.
Specifically, it removes the need for applications to use
platform-specific serverless API, which commonly differs between cloud
platforms. Further, it provides a unified way to express task-based workloads, 
to monitor the execution of tasks and to handle faults.

Underlying Pilot-Streaming is the pilot abstraction, which provides a unified
platform- and infrastructure-agnostic way to acquire resources and execute tasks on them. We extended Pilot-Streaming to support the allocation
of serverless resource containers. After the acquisition of the resources,
applications can manage their workload typically comprised of parallel and
dependent tasks on these resources~\cite{pstar12,saga_bigjob_condor_cloud}.
The tools support workloads comprising of different sets of dependent and
parallel tasks with different characteristics, such as task granularity,
compute, memory, and I/O demands. Data-parallel and streaming workloads
typically comprise of many short-running tasks that in the case of streaming
are generated in response to incoming data. \jhanote{Suggest consistency in IO
vs I/O. We use the latter in many places in this article.}\alnote{change to 
consistent use of I/O}

The pilot abstraction is exposed via the Pilot-API and consists of two
entities: \pilotjob  which represents a user-defined set of resources, and
\computeunit which is a task representing a self-contained set of operations
and is the key abstraction for expressing the application workload. Resources 
can be requested using the \pilotjob class. Once the \pilotjob has been 
instantiated,  \computeunits can be submitted to this instance.



Figure~\ref{fig:figures_pilotserverless} illustrates the different types of
infrastructures supported by the pilot abstraction. Initially,
pilot-jobs~\cite{pstar12} focused on the provisioning and management of HPC
and cloud resources (compute nodes, VMs), and the execution of the application
workload on these. Pilot-Data~\cite{pilot-data-jpdc-2014} and
Pilot-Streaming~\cite{pilot-streaming} extended the pilot abstraction to
externally provided distributed execution engines, e.\,g.,
Spark~\cite{Zaharia:2010:SCC:1863103.1863113} and Dask~\cite{dask}. This extension allowed the support of higher-level, data-intensive workloads that
rely on the capabilities of these execution engines. With serverless the
operational responsibility for the execution engine shifts from the application
to the infrastructure provider.

\begin{figure}[t]
  \centering
 \includegraphics[width=0.38\textwidth]{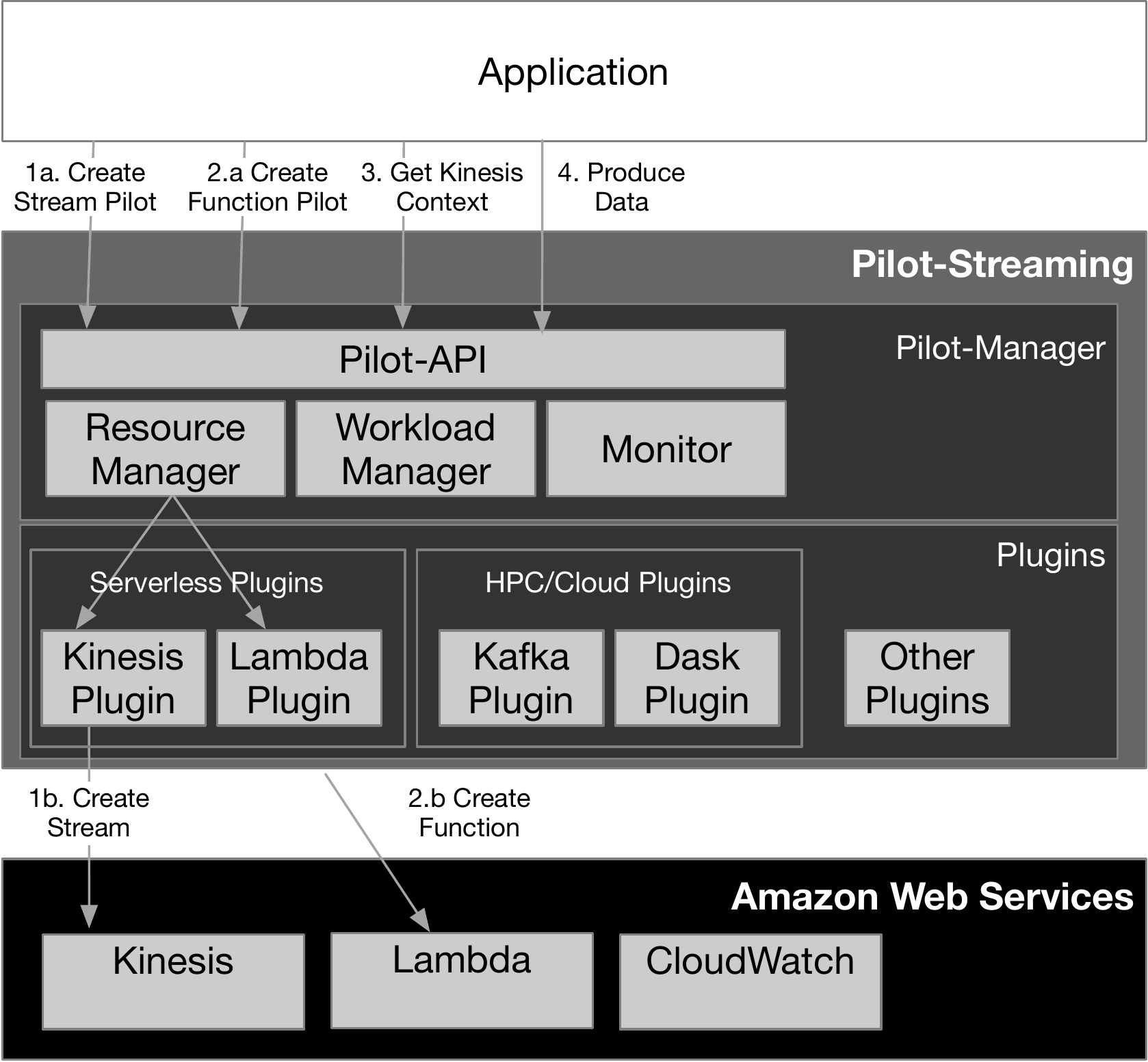}
  
\caption{\textbf{Pilot-Streaming on Serverless High-Level Architecture and Interactions:} Pilot-Streaming enables the unified management of batch and streaming compute tasks on serverless infrastructure. It orchestrates the setup of Kinesis and Lambda and manages the streaming workload. \label{fig:figures_pilotserverless_interactions}\upp\upp\upp}
\end{figure}



Pilot-Streaming is based on a modular architecture and provides several
plugins for supporting a diverse set of infrastructures and frameworks (see
Figure~\ref{fig:figures_pilotserverless_interactions}). The HPC and cloud
plugins~\cite{pilot-streaming} enable the usage of the pilot abstraction for
streaming applications on these platforms, i.\,e., it supports the deployment
and execution of Kafka and Dask on HPC and EC2 cloud instances. Using the plugin architecture, Pilot-Streaming was extended to support the serverless services AWS Lambda and Kinesis.  The Pilot-Manager continues to 
provide a unified interface -- the Pilot-API -- for running \computeunits on
these platforms, but also serves as an orchestrator for managing data and
compute across the different platforms.

\jhanote{Please tighten and clean up the previous paragraph. there is
conflation between pilot-framework and pilot-streaming. Also, is the later is
an abstraction and the former an implementation? If so, next paragraph needs
to read consistently with the above}\alnote{made more consistent. Only talking 
about Pilot-Streaming now. Pilot-Framework was still a leftover... There should 
not be a Pilot-Framework anymore in this paper. Here we only talk about the 
implementation, i.e. Pilot-Streaming}


In the following, we discuss the interactions between the components of
Pilot-Streaming. First, the application  typically initiates a Kinesis
\pilot and allocates a stream with a defined number of shards (see step 1a and b
in Figure~\ref{fig:figures_pilotserverless_interactions}). For this purpose,
the user needs to create a \pilotdescription, which provides a normative way to
specify resources for a streaming broker, e.\,g., the number of topic shards
for Kinesis and Kafka can be specified using the same attribute.
Pilot-Streaming then allocates resources for Kinesis using the 
platform-specific plugin, which encapsulates the necessary details. In cases 
where the
message broker is outside of the managerial scope of the application, this step
can be skipped and the producing, and the consuming application can directly 
connect to the broker.

\jhanote{Paragraph way too long. Break up in 3 maybe even 4 paragraphs}\alnote{shortened}

In step 2a and b, the Function \pilot for managing Lambda is created.
Pilot-Streaming supports the execution of arbitrary Python code as Lambda
function. Pilot-Streaming deploys the function on the serverless
environment and automatically handle serialization of the function and its
dependencies and the transfer to Lambda. Also, the usage of Lambda
Layers is supported. The \pilotdescription is used to specify and control the
parallelism in an infrastructure-agnostic way, while allowing the support for
infrastructure-specific capabilities, such as layers or memory limits on Lambda.

\jhanote{I'm very if not totally confused between pilot-streaming and
pilot-serverless.}\alnote{hopefully improved with the consistent removal of Pilot-Serverless in favor of Pilot-Streaming}

Pilot-Streaming supports two usage modes: (i) the submission of arbitrary
compute tasks to Lambda; (ii) the invoking of compute tasks in response to
incoming data events. For the first usage mode, the pilot abstraction can be
used to formulate common task-parallel patterns, e.\,g., for implementing data
parallelism or compose complex DAGs. For the second usage mode, the
\pilotdescription can be used to connect input data streams to the Lambda
function. Then, the framework will automatically setup this mapping. A task is
then automatically spawned in response to an event, e.\,g., a new message
arrived at the broker.

In summary, Pilot-Streaming simplifies the development, deployment, and
execution of serverless streaming applications. In addition to providing a
consistent way to allocate resources, it provides many functional
enhancements, such as the ability to express task-level parallelism more
efficiently than native FaaS APIs. The pilot abstraction provides a
well-designed and easy-to-use API for task-based workloads that can be
utilized to express different types of parallelism, from bag-of-tasks,
data-parallelism to streaming on heterogeneous infrastructure, making it
particularly well-suited for workloads required to run on a diverse set of
infrastructure. These capabilities provide the basis for achieving
high-performance and scalable applications, which we discuss in the next
section.

\section{StreamInsight: Performance Characterization and Modeling}
\label{sec:analytical}
\label{sec:perf_model}

\alnote{Look at parameter sensitivity, Uncertainty quantifications. Regime changes between different linear function }

\alnote{to incorporate:  For example, compute units need to be assigned to the
right size containers. As available resources per task is limited, this often
means that the computational and data requirements need to aligned to meet the
strict concerns of the serverless runtime. Other types of infrastructures, for
example message brokers like Kinesis, require the manual management of shards
and partitions. It is critical to allocate the right number of shards for a
pipeline as this controls the I/O and parallelism, and thus, has a significant
impact on the throughput. Over-provisioning results in unnecessary costs and
worse performance.}

Understanding the complex and dynamically varying performance characteristics
of streaming systems is necessary to ensure stable system execution by
allocating sufficient resources and configuring frameworks appropriately. The
performance depends on multiple variables, as the degree of parallelism and
the available memory. The number of attributes in the involved systems is high,
and often these attributes are not independent. Thus, performance data is 
challenging to collect, normalize, analyze, and compare. Furthermore, the 
combinatorial space
of parameters is ample, and thus, a careful selection of the most significant
factors to investigate is critical.


\emph{StreamInsight} supports the end-to-end process of performance experimentation and modeling of streaming applications and systems from experimental design, implementation, automation to analysis and modeling.  \jhanote{please check?}\alnote{fixed} It can be used to identify 
and quantify bottlenecks, test streaming systems under various traffic loads, fine-tune
system configurations, optimize resource allocations, and predict the
performance of applications. Further, the system can serve as a building block
for higher-level functionality,  such as predictive autoscaling.

StreamInsight comprises of two components: the \emph{Streaming Mini-App}
framework~\cite{pilot-streaming} that simplifies data collection and
automation of experiment, and a \emph{Universal Scalability Model
(USL)}~\cite{Gunther1993ASC} based performance modeling, analysis and
prediction framework. USL provides a well-established analytical model as a
foundation for understanding, explaining and predicting streaming systems. 
\jhanote{Whatkind of systems?}\alnote{fixed} The Streaming Mini-App 
framework~\cite{pilot-streaming} is used to simulate complex streaming
applications from data production, brokering to processing. The capabilities of
the Streaming Mini-App framework allow application developers to automate the
exploration of large parameter spaces of different frameworks and application
configurations.

The Mini-App framework is built on Pilot-Streaming, which is used to manage
resources, stream processing runtimes, and the message broker. We extended the
Mini-App framework to support the collection of comprehensive performance data
on serverless infrastructure in addition to HPC infrastructure and provide the
ability to trace data across all components, i.\,e., data source, brokering
and processing system. For this purpose, the framework assigns a unique run
id, which is propagated to all involved components. This way events can be
attributed to a specific benchmark runs.

The instrumentation system is architected in a modular way allowing the
developer to easily add/remove metrics for all components. Currently, data can
be collected from Pilot-Streaming, the message brokers, processing frameworks
and cloud-based log services, such as AWS CloudWatch. The framework is
architected in a modular way, i.\,e., the data collector can easily be extended
to support new systems and infrastructure. It can simulate different data rates
and characteristics (e.\,g., message sizes) aiding the collection of
performance data. To conduct measurements at the maximum sustained throughput,
the framework utilizes an intelligent backoff strategy during data production.

\jhanote{Would it make sense to move the detailed discussion of Mini-App to
Section 2 --- Background and related work?}\alnote{I am kind of hesitant as the Mini-Apps are a contribution. The efforts for adding serverless instrumentation is not negligible}

\alnote{Move somewhere: While HPC and serverless differ in the provided
abstractions and management, they share many common characteristics with
respect to performance and scalability.  For example, both Kinesis/Lambda and
Kafka/Dask implement data parallelism based on data partitions; Kinesis refers
to a partition as shard. The main difference between Kinesis and Kafka is that
Kafka requires a careful consideration of storage, node type and number of
nodes for scaling. Kinesis provides a defined QoS for streaming: 1 MB/sec per
shard. Details of the underlying infrastructure are not known and required. In
the following, we propose a unifying model for both HPC and serverless.}


\jhanote{Need clarity on StreamInsight. Distinguish what it is, what it is
used for, and how it achieves it? StreamInsight is not the performance model
itself, correct? We mention, ``It enables design experimentation, automation,
analysis and modeling of application and system performance'' but (i) isn't
analysis and modeling of application and system performance, the role of the
performance model too? (ii) do we come back  and discuss automation and
analysis?}\alnote{added: ``StreamInsight comprises of two components: the \emph{Streaming Mini-App}
framework~\cite{pilot-streaming} that simplifies data collection and automation
of experiment, and an \emph{Universal Scalability Model
(USL)}~\cite{Gunther1993ASC} based performance modeling, analysis and
prediction framework.'' Will come back to this in conclusion section }

\subsection{Performance Model} 
\label{perf_model}


We build a model for understanding and predicting the throughput of a streaming
system. We begin with a definition of the relevant variables. A streaming
system consists of the message broker ($br$), and the processing system ($px$).
The performance is measured by two metrics: the latency ($L$) and throughput
($T$). $L$ is defined as the time difference between two states, e.\,g.,
$L^{px}$ is the time between arrival and processing of message in the
processing system. $L^{br}$ is the time between message production and
its availability at the broker.


Throughput $T$ is defined as the number of events (e.\,g., messages) a system
handles in a certain amount of time. The throughput of the stream processing
system ($T^{px}$) depends upon the parallelism and the number of partitions
$N^{px}(p)$ and nodes $N^{px}(n)$. We measure and analyze the maximum
sustained throughput, i.\,e., the optimal load a streaming system can handle
without performance deterioration (e.\,g., due to
back-pressure)~\cite{DBLP:journals/corr/abs-1802-08496}. Further control
parameters are the machine $M$, the message size $MS$, and the workload complexity $WC$. Table~\ref{tab:modelparam} summarizes the parameters of the
model.\jhanote{We might want to simplify given the discussion above.}
\jhanote{I'm not sure I understand why $WC$ is a variable. Its an attribute or
a property of the model.}\alnote{The model is in this case not USL but the K-Means model. I replace that with workload complexity. Is that better?}


\note{separate model from experimental setup. If they are parameters then you 
should not constrain them. They are free. move the following paragraph to 
experiments section}

\note{do not go from general to specific again. move to opening}
\jhanote{Do you mean: ... ``$T$'' of
a streaming systems and its dependence on $N(p)$?}\alnote{yes, I meant that. fixed} 




According to USL, $T(N)$ of a system is represented as follows:
\begin{equation*}
\footnotesize
\begin{split}
T(N) \ = \ \frac{N}{1+\sigma(N-1)+	\kappa N(N-1)}
\end{split}
\label{usl}
\end{equation*}

For stream processing systems, $T =T^{px}$ and $N = N^{px}(p)$. The expression
has two parameters: $\sigma$, which measures the serial overhead in the
system, and $\kappa$, which captures the coherence between all processors. For
example, $\sigma$ can be used to quantify overheads, such as serialization;
$\kappa$ measures the all-to-all communication typically required, e.\,g., for
sharing model parameters across all tasks. A $\sigma$ and $\kappa$ of 0
indicate optimal scalability. The larger both terms are, the worse the
scalability. USL is well suited for modeling streaming systems and provides the right level of granularity while maintaining a high degree of interpretability.
Further, USL does not require low-level timing to model the system.
Gunther~\cite{DBLP:conf/pdcs/Gunther05} showed that USL generalizes Amdahl's
laws~\cite{Amdahl:1967:VSP:1465482.1465560} and adds meaningful extensions,
e.\,g., to explain performance degradations.



The different components of the physical system have a different impact on the
model parameters. Due to the higher parallelism, larger $N^{px}(p)$ and
$N^{px}(n)$ create more contention on shared resources, such as shared
filesystems and networks. The ratio of $N^{px}(p)$ to $N^{px}(n)$ influences
both $\sigma$ and $\kappa$ as it determines the amount of network traffic
required to communicate between the partitions.

\jhanote{the following paragraph needs to be either removed or significantly
worked. It is not clear what ``model'' is, and therefore model size or model
complexity. Thus, I'm commenting it out for the time} \alnote{rename MC to WC (workload complexity) - I am fine with removing this as we have a similar analysis later}

\jhanote{BTW, $N(p)$ and $N(n)$ aren't defined.}\alnote{added superscripts}


\begin{table}[t]
	\scriptsize
	\centering
	\begin{tabular}{|p{1.5cm}|p{4.5cm}|}
	\hline
	$L$  &Overall Latency \\ \hline
	$L^{px}$ &Latency Processing System \\ \hline
	$L^{br}$ &Latency Broker System\\ \hline
	$T$  &Overall Throughput \\ \hline
	$T^{px}$ &Throughput Processing System \\ \hline
	$T^{br}$ &Throughput Broker System\\ \hline \hline
	$N^{px}(n)$ &Number Nodes Processing System \\ \hline
	$N^{px}(p)$ &Number Partitions Processing System \\ \hline
	$N^{br}(n)$ &Number Nodes Broker System \\ \hline
	$N^{br}(p)$ &Number Partitions Broker System \\ \hline\hline
	$M$  & Machine and Infrastructure\\\hline
	$WC$ & Workload Complexity\\\hline
	$MS$ & Message Size\\\hline
	\end{tabular}
	\caption{Model Dependent, Independent and Control 
	Variables\label{tab:modelparam}\upp\upp}
\end{table}


We assume a stable system operating at its maximum sustained throughput $T$,
which is ensured by using an intelligent backoff strategy during data
production. Further, overheads for initialization of the streaming application
and infrastructures, i.\,e., the time for starting the streaming framework via
the Pilot-Streaming, are not considered. We fit the data to USL using a
non-linear regression model provided by the USL R package~\cite{uslr}.

\subsection{Experiments and Characterization}
\label{sec:character}

We characterize the performance \jhanote{``on'' or ``of'' ?}\alnote{of?} of different infrastructures ($M$): (i) Kinesis/Lambda serverless processing on AWS, and
(ii) Kafka/Dask stream processing on Wrangler and Stampede2 (both HPC). In
both scenarios, we deploy a Streaming Mini-App framework, i.\,e., the
synthetic data generator, \jhanote{relate back to mini-app? e.g., of the
mini-app}\alnote{improved} the broker, and distributed execution engine on  HPC, respectively AWS. For HPC, we use the XSEDE machines Stampede2 and Wrangler. Each Wrangler
node provides 48 cores and 128\,GB RAM. On Stampede2, the Knights Landing
nodes with 68 cores, 96\,GB RAM were used. For serverless, we use AWS Lambda
and Kinesis with different container sizes and numbers of partitions. For the
synthetic data generator, we utilize \texttt{m5.4xlarge} nodes with 16\,cores
and 64\,GB memory.

We investigate the clustering algorithm K-Means~\cite{kmeans} as a
representative workload. K-Means is well understood and commonly used in
streaming applications to detect abnormal behavior. K-Means has a complexity
of $O(nc)$ with $n$ being the number of points and $c$ the number of cluster
centroids. The algorithm comprises of two phases: First, K-Means computes the
Euclidean distance between all points ($n$) and the centroids ($c$), which
corresponds to a complexity of $O(nc)$. Then, the new centroid positions are
computed by averaging the positions of all points assigned to a centroid. 

We use the MiniBatch K-Means of scikit-learn~\cite{kmeans-scikit} and update
the K-Means model continuously based on the incoming data. The model is shared
across tasks using file storage (S3 on AWS, Lustre filesystem on HPC). We vary
the workload complexity $WC$ by using different numbers of centroids (between 128 and 8,192): the higher the number of centroids the more compute is necessary,
\jhanote{We should be careful to not overload model in Section IV-A. The reader
will think that ``model'' in \S IV-A is for performance model. But given the
remark of above, it appears $WC$ is unrelated to the performance 
model?}\alnote{I agree. renamed model complexity to workload complexity} and
the I/O for writing/reading the model is increased. Further, we evaluate
different message sizes $MS$: 296\,kb for 8,000 points, 592\,kb for 16,000
points and 962\,kb for 26,000 points.

\alnote{refine. add values for Wrangler and Stampede}
For this experiment, we constrain the variables as follows: To
meet the memory requirements of the application, we select an appropriate
core/node ratio: $N^{px}(n)$/$N^{px}(p)$. On both Wrangler and Stampede2, we 
use 12 cores/nodes, which corresponds to 11\,GB per core on Wrangler and 8\,GB per core on Stampede2. Further, we use the same number of brokers and processing
nodes: $N^{px}(n) = N^{br}(n)$.



\subsubsection{Lambda Memory}
 
\begin{figure}
\centering
\includegraphics[width=.37\textwidth]{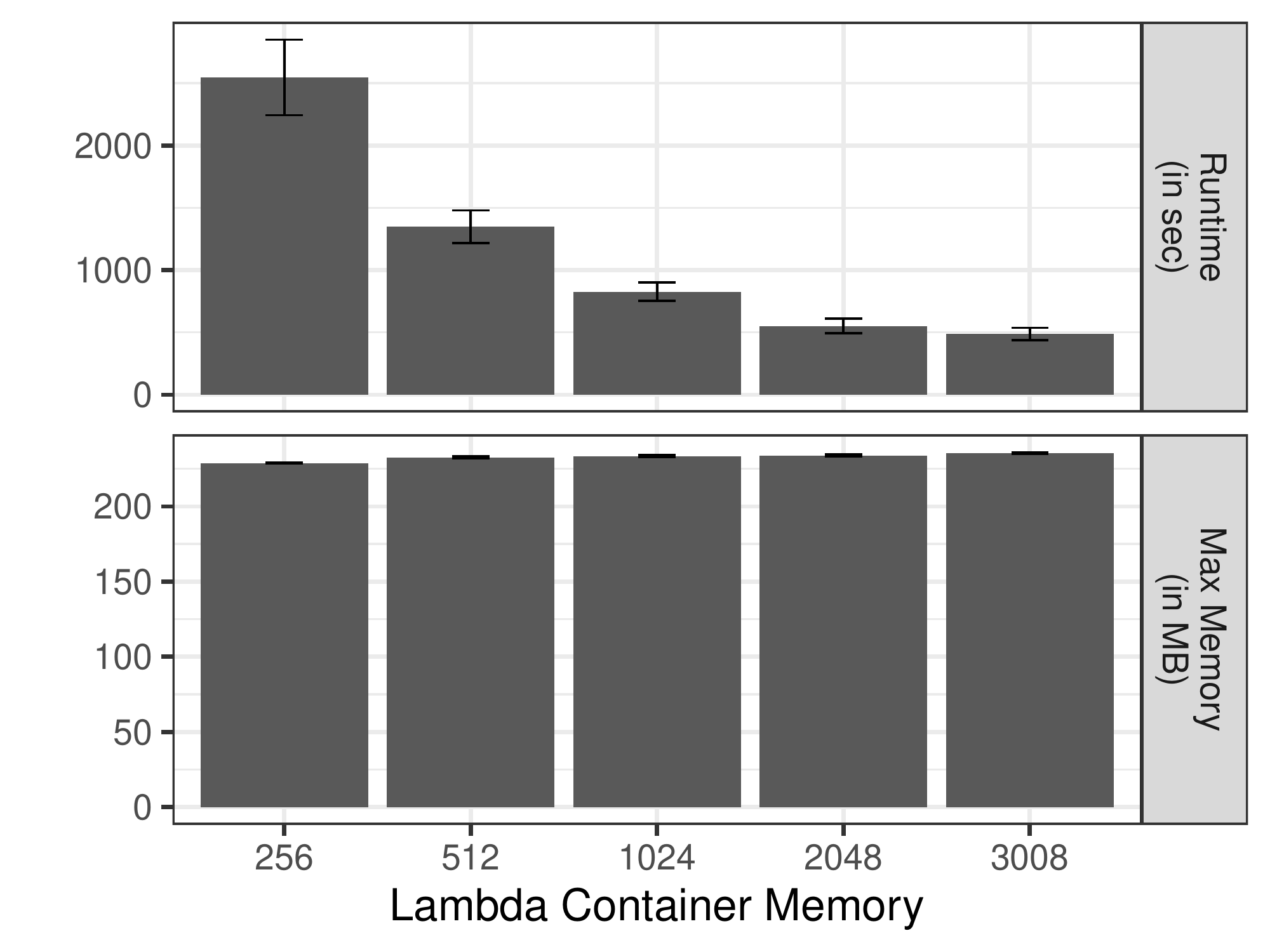}
  \caption{\textbf{Lambda Container Memory for 8,000 points and 1,024 centroids:} Lambda containers with a larger 
  amount of memory provide more compute capacity and thus, enable shorter 
  runtimes. The fluctuation in the data is significantly lower for larger 
  container sizes. \alnote{add baseline performance on Wrangler/Stampede}\upp\upp}
  \label{fig:lambda_runtime}
\end{figure}

An important parameter for \computeunits executed within Lambda  
\jhanote{What is a Lambda application?}\alnote{rephrased using our Pilot-Abstraction vocab} is the amount of memory.
Figure~\ref{fig:lambda_runtime} illustrates the runtime of the K-Means Lambda
function broken down by the requested lambda container memory. We investigate
the memory usage for a message size of 8K points and 1024 clusters. Even
though the actually used memory for the lambda function nearly remains
constant, the runtime decreases with larger memory runtimes (currently max
3,008 MB) indicating that AWS scales the CPU allotment proportional to the
memory.

\subsubsection{Lambda vs. Dask}

\begin{figure}[t]
  \centering
\includegraphics[width=.45\textwidth]{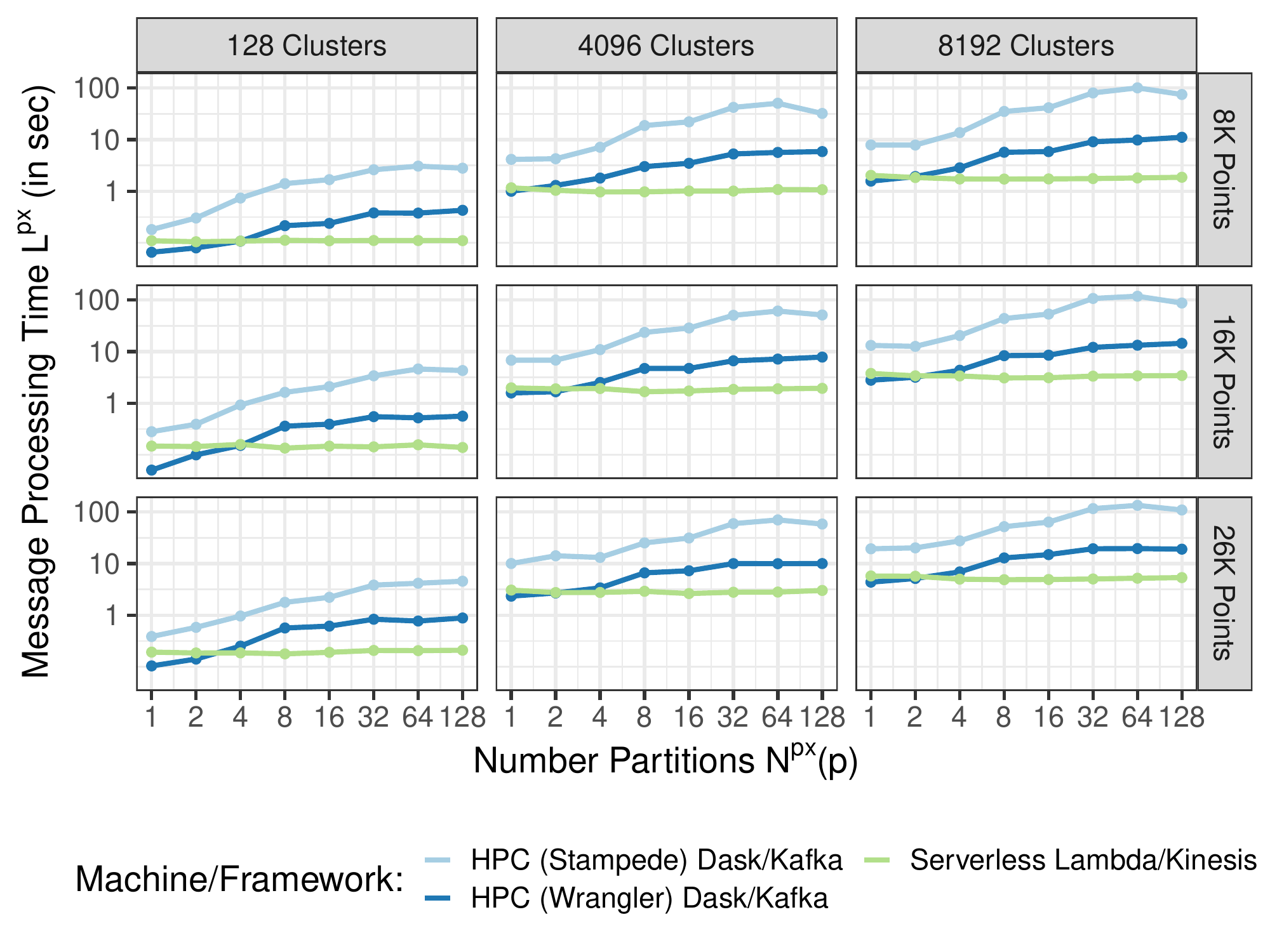}
  \caption{\textbf{Message Processing Time $L^{px}$ for K-Means on AWS Lambda and HPC:} 
  Broken down by number partitions, message size, and workload complexity. The 
  processing times increase with the message size and 
  a higher number of clusters. While for Lambda the processing times remain 
  constant with increasing parallelism, we observe a negative impact for 
  Dask/Kafka on HPC  due to the use of shared filesystem and network resources.
  \label{fig:figures_streaming_processing_time}\upp\upp}
\end{figure}


Figure~\ref{fig:figures_streaming_processing_time} investigates the processing
latency per message $L^{px}$ for different numbers of messages and centroids.
For Lambda, we use 3\,GB RAM containers. The processing latencies increase with
the number of points and centroids for both Dask and Lambda due to the increase
of the computational, I/O and memory demands of the processing function. For
Lambda, the processing times remain stable with higher parallelisms, i.\,e.,
higher partition counts. The number of Lambda containers is managed 
completely by AWS. AWS  never starts more containers than Kinesis
partitions. During all experiments, we observed at most 30 concurrent 
containers.

\begin{figure}[t]
  \centering 	\includegraphics[width=.45\textwidth]{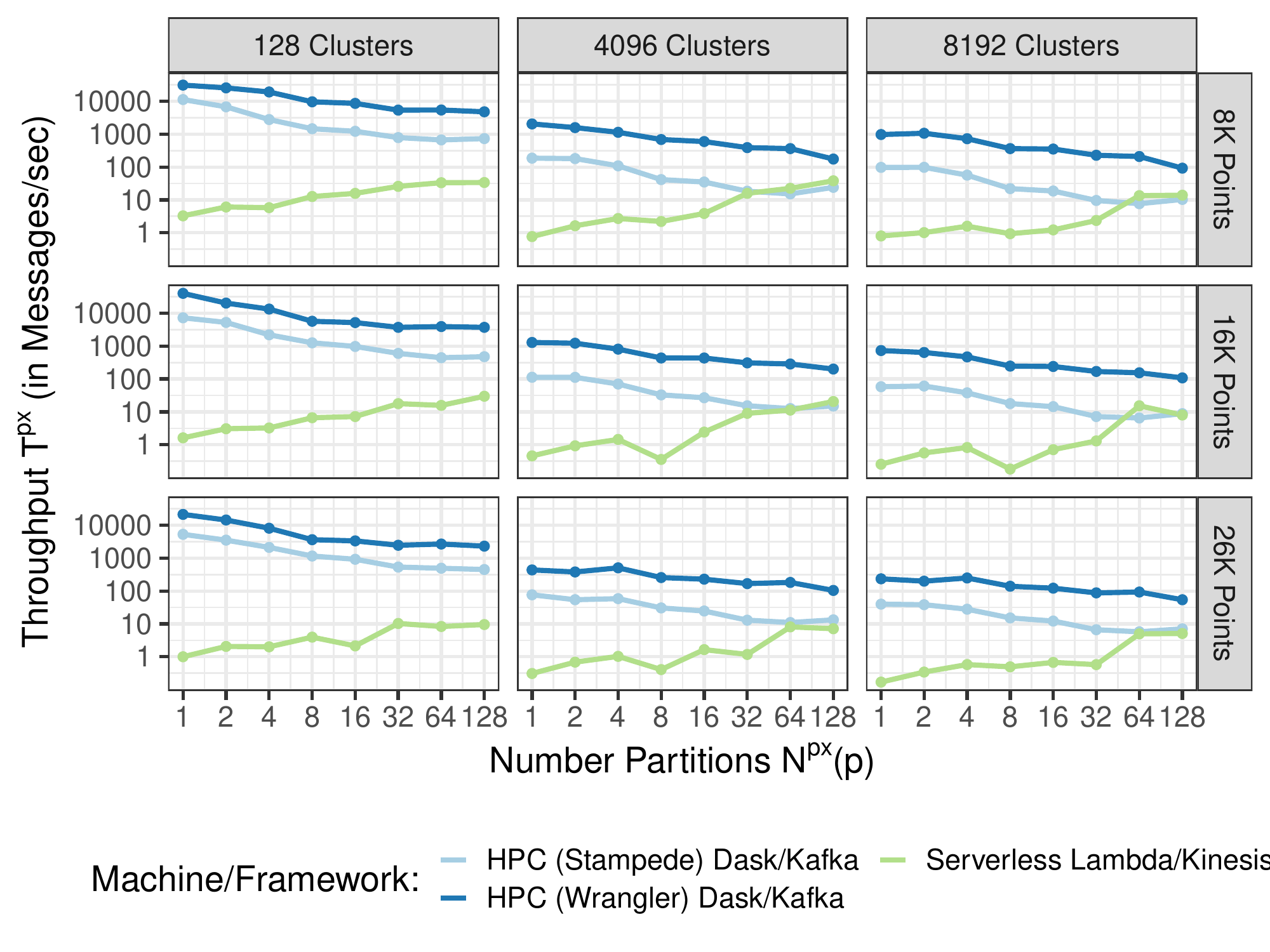}
  \caption{\textbf{Lambda and Dask Throughput $T^{px}$ for K-Means on AWS Lambda
and HPC:} The increased processing times also impact the throughput and
speedup. For scenarios with higher compute to I/O ratio a small speedup is
observable for Dask until 4 partitions.
\label{fig:experiments_kmeans_dask_throughput_speedup}}

\end{figure}

While Kinesis/Lambda stream processing provides a predictable performance, for
Dask $L^{px}$ increases with the number of partitions indicating system
contentions. This also results in a degradation of the throughput for larger
$N^{px}(p)$ as shown in
Figure~\ref{fig:experiments_kmeans_dask_throughput_speedup}. For the more
compute-intensive scenarios, i.\,e. in particular larger model sizes such as
8,192 clusters, a small speedup of up to 1.2 is observable for Dask/Kafka until
4 partitions. In the following, we analyze the data using USL.

\alnote{read/write concurrency limitations... as all producers and consumers
are on the same network and filesystems}

\subsection{Model Analysis} 
\label{sec:perfmodel}
\alnote{Extension of model: model impact of reconfiguration (required 
reconfiguration time, improved throughput}

\begin{figure*}[t]
\centering
\includegraphics[width=.8\textwidth]{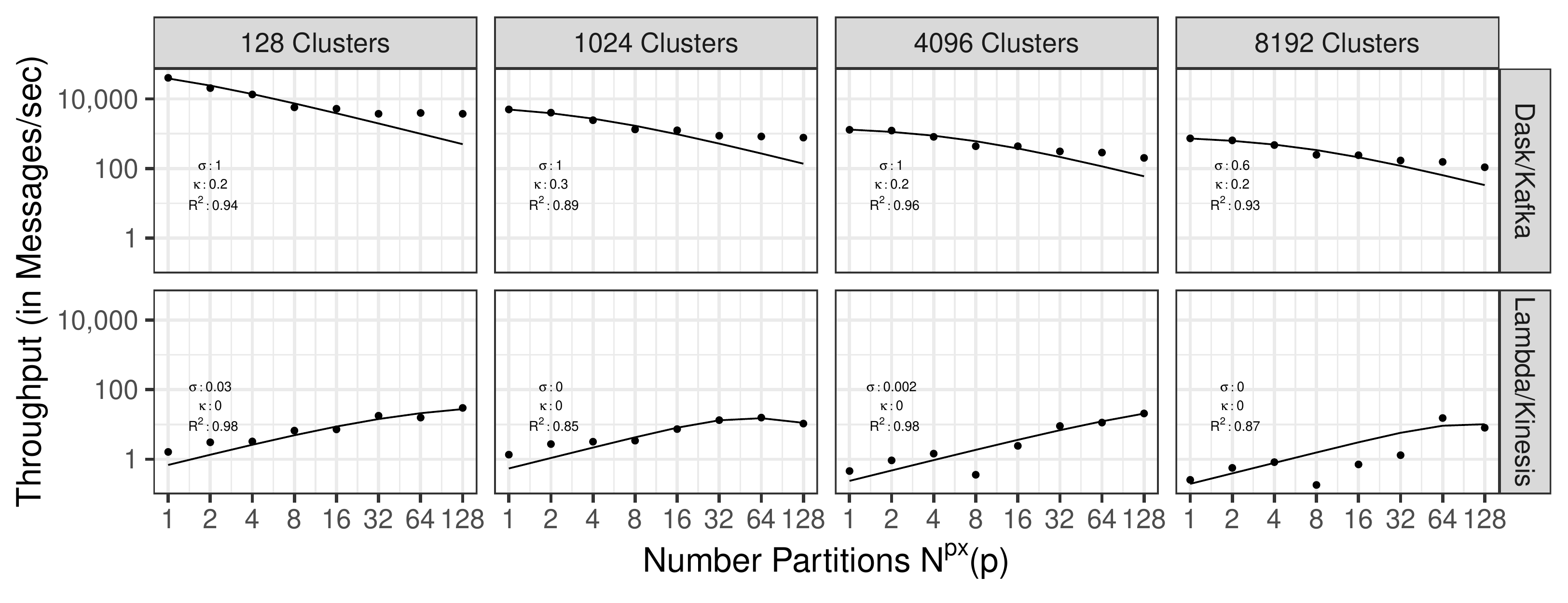}
\caption{\textbf{Model Fit Universal Scalability Law applied to Lambda and
Dask Streaming (Wrangler):} \jhanote{the fontsize of $\sigma$ and $\kappa$ are too small}\alnote{increased as much as I could}
The figure shows how the USL model was fitted to different scenarios. In all
scenarios, the message size is 16,000 points. USL is suitable to describe the
performance of streaming applications and enables StreamInsight to quantify
overheads. For Kinesis/Lambda, USL produces a very small  $\sigma$ and $\kappa$ 
explaining the optimal scalability. For Kafka/Dask, we observed larger 
coefficients explaining the severe performance degradation. \jhanote{What do the 
fitted values of $\sigma$ and $\kappa$ represent / mean?}\alnote{explained}\upp 
}
 \label{fig:experiments_usl_fit}
\end{figure*}

Figure~\ref{fig:experiments_usl_fit} shows the throughput and the computed USL
model for the different variables $M$, $WC$, and $MS$. The K-Means application
is loosely coupled; all tasks process a partition independently with a
minor synchronization for model updates. Further, there are shared I/O
resources used by the broker and the processing engine in particular on HPC. These characteristics are reflected in the $\sigma$ and $\kappa$ coefficients of the USL model. $\sigma$ quantifies the serial overhead, while $\kappa$ measures the all-to-all
communication between all nodes. All-to-all communication can e.\,g. be
required to update the model parameters between all tasks. We also noticed on
both HPC environments severe performance degradation for larger $N$ due to I/O
issues caused by the shared filesystem.


The scalability depends significantly on the infrastructure. While HPC provides 
a better absolute performance, cloud infrastructures are more predictable. For
HPC the peak performance is already reached in many cases using a single
partition. The system performance degrades with increased parallelisms due to
contention and coherencies overheads. Running both data production, brokering,
and processing (including complex coordination for sharing model parameters) on
the shared filesystem is the likely cause. Lambda and Kinesis provide better
resource isolation.



For Lambda, the throughput increases with the number of partitions thanks to
the greater parallelism. Smaller message and model sizes have a higher
throughput. USL produces a contention coefficient $\sigma$ and a coherence
coefficient $\kappa$ of close to zero explaining almost optimal scalability.
These small coefficients shows that Lambda containers are well isolated, providing a predictable performance enabling AWS to provide predictable SLAs.

For Dask, a $\sigma$ between 0.6 and 1 explains that 
the scalability bottleneck is caused by resource contentions, e.\,g., through 
shared memory, network and filesystem. However, $\kappa$ indicates that some
significant coherency due to cross-communication between all processors,
e.\,g., the synchronization of the model updates via the shared filesystem.
Thus, the peak scalability of the system is already reached with a single
partition. Further, it must be noted that Dask/Kafka was carefully fine-tuned
to the HPC machine by carefully choosing the right memory/core ratio and the 
location of the Kafka data log files. However, the number of shared resources 
that impact performance is significantly larger on HPC than on serverless.

\subsection{Model Evaluation}
\label{sec:modeleval}

\begin{figure}[t]
  \centering
    \includegraphics[width=.4\textwidth]{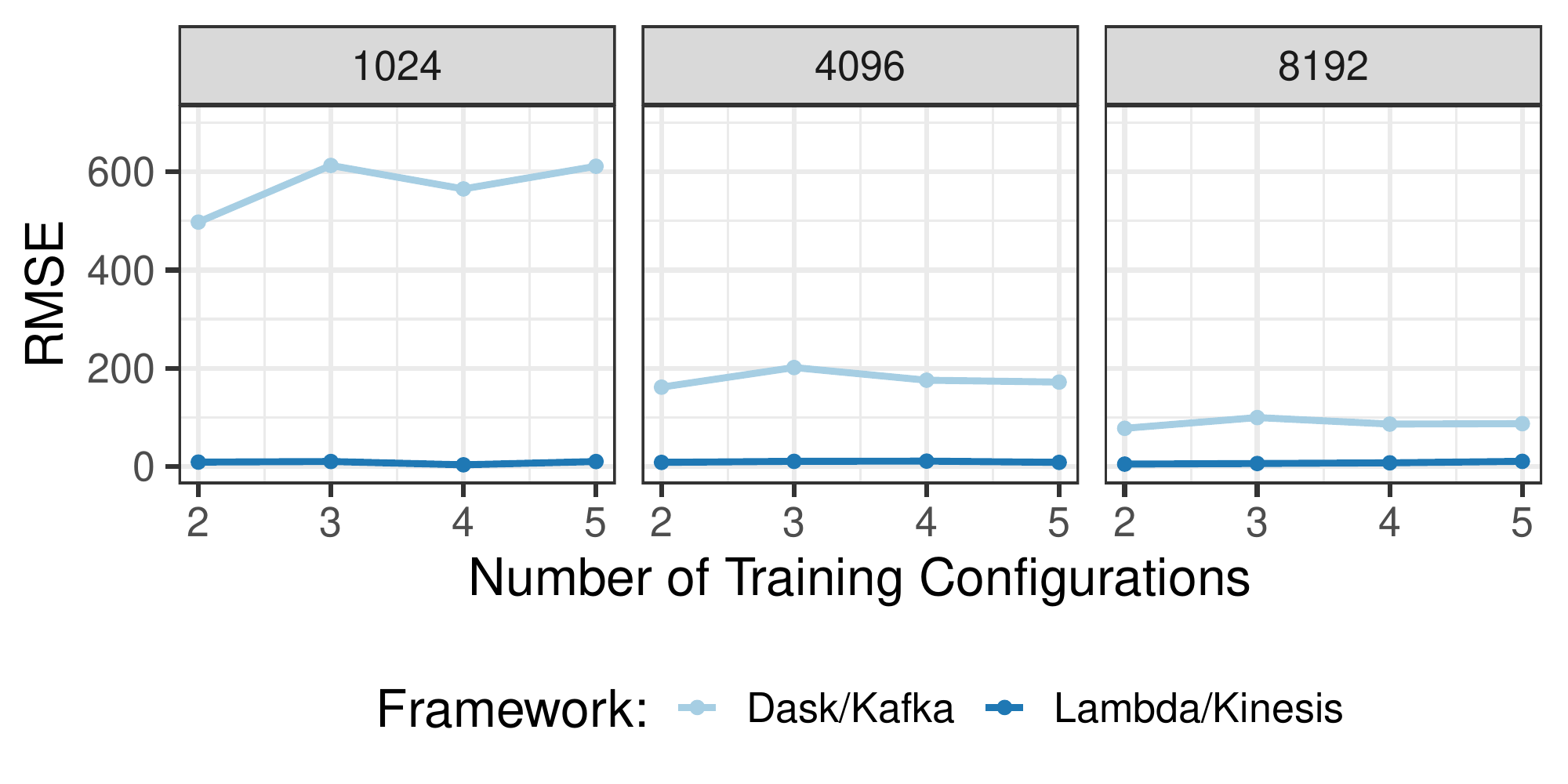}
  \caption{\textbf{RSME for Different Sizes of Training Data:} A small number of observations is sufficient to obtain a well-fitted model.\upp}
  \label{fig:experiments_05.02_Fig11f_Dask_fig17_rmse}
\end{figure}

For evaluation, we investigate the model fit and the performance
of the model on unseen data. We observed a training $R^{2}$ between 0.85 and
0.98, indicating that the model can capture a considerable proportion of the
variance within the data. For unseen data, we split the benchmark data into a
training and test set. We utilize a different number of training configurations
to create a performance model. We investigate the root mean squared error of
the predictions on the unseen test data of the remaining configurations.
Figure~\ref{fig:experiments_05.02_Fig11f_Dask_fig17_rmse} illustrates the
results. 

A small number of observations, i.\,e., 2-3 training configurations are enough
to create a well-performing model. Due to the small amount of data, it can
easily be used to identify optimal configurations for production systems. In
general, the Lambda/Kinesis is more predictable than the Dask/Kafka model. For
Dask, we observe a higher RSME for short-running tasks, i.\,e., smaller message
and model sizes. For these configurations, the contention and coherence caused
by the shared resources are higher, making the prediction is less precise.


\alnote{Maybe integrate into conclusion: In summary, StreamInsight allows the quantification of the
overheads and scalability for streaming applications and infrastructure using
only very few data points. Due to the small amount of data it can easily be
used to identify optimal configurations for production systems. In the future,
we will use the model to dynamically scale streaming infrastructures. The
performance models allow us to explain and quantify properties, such
as linear scalability, diminishing returns, bottleneck limitations, but also
performance degradations. }

\section{Conclusion and Future Work}
\label{sec:conclusion}

\jhanote{Should not forget to connect to HPC application scenarios / motivation /
requirements. Currently missing} 

\note{Dan Reed: Compute Continuum - \url{https://www.hpcdan.org/reeds_ruminations/technology/}}

\note{What are the types of Experiments in the loop? Control loop.
Computing Loop is influencing the experiment
Experiments that influence compute loop. The typical IoT approach is that data is funneled into computing. Instantiation of streaming concept
Computing in experimental loop is a bit more exciting. Issue of scale? More challenges.
wine case study: improve placement, sensitive, data placement - you have more 
1-2 paragraphs in introduction to unpack xloop
}


We presented the Pilot-Streaming on serverless and StreamInsight  designed to
support the development, deployment, and execution of streaming applications.
Pilot-Streaming on serverless demonstrated that the pilot abstraction's task
model is well suited to express and manage serverless streaming workloads on
serverless and HPC. The pilot abstraction enhances the
expressiveness of serverless abstractions, e.\,g., by adding the ability to
coordinate across multiple tasks required for many applications, such as
machine learning. Further, Pilot-Streaming enables the interoperable use
of serverless, cloud, and HPC.

StreamInsight provides comprehensive tools for creating performance experiments
and analyze the data. We evaluated StreamInsight using different complex
machine learning tasks and showed that the USL approach is well suited to predict
the scaling properties of streaming applications requiring only small amounts
of data. USL allows the quantification of properties, such as diminishing
returns and communication bottlenecks. We found that using Kafka/Dask based
stream processing on HPC provides better absolute performance than
Kinesis/Lambda. However, the performance on HPC degrades when scaling
due to I/O issues quickly. Kinesis/Lambda provide better scalability and
predictability. Serverless is, however, subject to several limitations, e.\,g.,
the strict walltime of 15 minutes. Also, currently no GPUs are supported. Thus,
for high-end workloads that require complex and large dependencies, e.\,g.,
TomoPy for light sources sciences or deep learning applications, serverless is
not suitable.

We will extend StreamInsight to support advanced
experimentation-in-the-loop-computing scenarios making use of complex edge,
fog and cloud infrastructure. We  will enhance Pilot-Streaming to support FaaS
infrastructures, in particular on edge and fog environments. With Greengrass,
AWS supports the execution of Lambda functions on the edge. By moving
serverless functions to the edge and thus, closer to the data, further
optimizations are possible. We will integrate StreamInsight into the resource
management algorithm of Pilot-Streaming so as to support predictive scaling,
viz., the ability to adapt the resource allocations and configurations to
changes in the incoming data rate(s). This will also enable the determination
of the amount of throttling of data sources to guarantee processing.

The above represent some advanced resource management capabilities that
emerging EILC scenarios will require. Pilot-Streaming and StreamInsight
provide dynamic resource management and a predictive resource allocation ---
two initial but instrumental building blocks towards these capabilities.

\alnote{StreamInsight and the pilot abstraction. provide two instrumental building blocks for implementing adaptive
infrastructure capabilities, such as predictive auto-scaling: The
pilot abstraction enables the dynamic resource allocation and StreamInsight
provides the ability to predict and thus, control resource needs
appropriately.
}

{\scriptsize{\it Acknowledgements: } We acknowledge support from  NSF DIBBS
1443054 and NSF CAREER OAC 1253644. XSEDE computational resources were made
available via XRAC allocation TG-MCB090174.}

\bibliographystyle{unsrt}
\bibliography{benchmark,saga,saga-related,pilotjob,radical_rutgers,radical_publications,bigdata,local,streaming}

\end{document}